\documentclass[preprint,12pt,authoryear]{elsarticle}

\usepackage[T1]{fontenc}
\usepackage[utf8]{inputenc}
\usepackage{natbib}
\usepackage{hyperref}
\usepackage{graphicx}
\usepackage{amssymb, amsmath, amsthm}
\usepackage{amsfonts}
\usepackage{epstopdf}
\usepackage{genmpage}
\usepackage{booktabs}
\usepackage{bookmark}
\usepackage{pdfpages}

\hypersetup{
    pdftitle={Bayesian between-subjects circular data analysis},
    pdfauthor={Kees Tim Mulder},
    pdfsubject={Extending Bayesian analysis of circular data to comparison of multiple groups},
    pdfkeywords={bayesian, anova, von mises, between-subjects},   
    bookmarksnumbered=true,     
    bookmarksopen=true,         
    bookmarksopenlevel=1,       
    colorlinks=true,
    citecolor=blue,
    linkcolor=red,
    urlcolor=blue,                 
    pdfstartview=Fit,           
    pdfpagemode=UseOutlines
}

\journal{Journal of Statistical Planning and Inference}

\begin{document}

\renewcommand{\thefootnote}{\fnsymbol{footnote}}

\begin{frontmatter}



\title{{Extending Bayesian analysis of circular data to comparison of multiple groups }}


 \author[label1]{K. T. Mulder\footnote{Corresponding author.}}
 \author[label1]{I. Klugkist}
 \address[label1]{Utrecht University, Department of Methodology and Statistics, Utrecht, The Netherlands}


\begin{abstract}
Circular data are data measured in angles and occur in a variety of scientific disciplines. Bayesian methods promise to allow for flexible analysis of circular data. Three existing MCMC methods (Gibbs, Metropolis-Hastings, and Rejection) for a single group of circular data were extended to be used in a between-subjects design, providing a novel procedure to compare groups of circular data. Investigating the performance of the methods by simulation study, all methods were found to overestimate the concentration parameter of the posterior, while coverage was reasonable. The rejection sampler performed best. In future research, the MCMC method may be extended to include covariates, or a within-subjects design. 

\end{abstract}


\end{frontmatter}



\section{Introduction}

Circular data are data measured in angles or orientations in two-dimensional space. For example, one may imagine directions on a compass ($0^\circ - 360^\circ$), times of the day ($0 - 24$ hours), or directions on a circumplex model, such as Leary's Circle \citep{Leary1957}. Circular data are frequently encountered in many scientific disciplines, such as biology, social sciences, meteorology, astronomy, earth sciences, and medicine.

The analysis of circular data requires special directional statistical methods due to the periodicity of the sample space. For example, two angles of $10^\circ$ and $350^\circ$ differ by only $20^\circ$, while if treated linearly the distance between them would seem to be $340^\circ$. A similar mismatch occurs for the arithmetic mean of $10^\circ$ and $350^\circ$, which is $180^\circ$,  while their correct circular mean is $0^\circ$. 

Three different approaches for analysis of circular data are discussed in the literature: the \textit{intrinsic} approach, which uses the von Mises distribution \citep{von1918ganzzahligkeit, damien1999fullbayes}; the \textit{embedding} approach, which employs the Projected Normal distribution \citep{Nunez-Antonio2005}; and the \textit{wrapping} approach, where distributions on the real line are wrapped around the circle \citep{ferrari2009wrapping}. The intrinsic approach is the most prominent in the literature, perhaps because this is currently the only approach which allows calculation of maximum likelihood estimates \citep{ferrari2009wrapping}. Additionally, mapping the circular sample space to a sample space in either $\mathbb{R}^1$ (wrapping) or $\mathbb{R}^2$ (embedding) generally leads to an increase in the amount of parameters to be estimated, which may make these methods more complex. Because of these reasons, the scope is limited to the intrinsic approach here. 

Due to the difficulty of working with a circular sample space, few methods have been developed in the field of analysis of circular data. An overview of available frequentist methods for analysis of circular data can be found in  \citet{fisher1995statistical} and \citet{mardia1999directional}. Bayesian methods offer a promising new approach not only in the field of statistics at large, but also specifically in the analysis of circular data. Main advantages of the Bayesian approach are the flexibility of Markov chain Monte Carlo (MCMC) methods used in Bayesian analysis, the lack of asymptotic assumptions, and the possibility to incorporate knowledge from previous research. Some work has been done performing Bayesian estimation on circular data without utilising MCMC methods \citep{dowe1996bayesian}, but such methods only perform point estimation without providing standard errors, while researchers are often interested in drawing inference.

In the case of directional statistics, MCMC methods may prove to be a flexible solution to the difficulty of drawing inference from circular data. A limited number of MCMC methods for circular data have been developed. Available methods generally employ the von Mises distribution, which is the natural analogue of the normal distribution on the circle. Early work by \citet{damien1999fullbayes} provided a Gibbs sampler for a single group by adding latent variables to the model. Metropolis-Hastings algorithms have been developed for circular distributions in general \citep{Bhattacharya2009} and for the von Mises-Fisher distribution, which is the generalization of the von Mises distribution to the sphere \citep{nunez2005bayesian}. Recent work has attempted to tune the parameters of a rejection sampling algorithm in order to obtain a computationally fast method to sample from the posterior of a von Mises distribution \citep{forbes2014fast}. Although different in approach, these methods have in common that they draw from the posterior of the von Mises distribution given one group of circular data, which can be used to describe properties of a single sample. None of the methods may be used to compare groups. 

In this paper existing MCMC methods will be extended to analyse data from between-subjects designs, where the research goal is to compare mean directions of multiple groups on a circular outcome. Many tests in between-subjects designs, such as ANOVA, assume equal variance across groups. Circular ANOVA methods that have been developed in a frequentist framework also carry this assumption \citep{harrison1988development, harrison1986analysis}. A main aim of this paper is thus to extend available MCMC methods to between-subjects designs, so that the method samples multiple mean directions and a single measure of dispersion. Then, the performance of these methods will be assessed to decide which is the most commendable. 

Section \ref{tf} provides the theoretical framework and notation for the von Mises distribution. Then, in Section \ref{methods}, three MCMC methods for between-subjects designs are discussed. These are compared by means of a simulation study in Section \ref{simstud}. Concluding remarks are given in Section \ref{discussion}.

\section{The intrinsic approach \label{tf}}

The MCMC methods discussed in this paper all fall within the intrinsic approach, where it is assumed that the data follow the von Mises distribution. This section will discuss basic properties of the von Mises distribution and provide a framework for the MCMC methods that will be discussed in Section \ref{methods}. The first four sections will be restricted to the von Mises distribution for a single group, while Section \ref{multiple} will introduce properties and notation to be used in the case with multiple groups. 

\subsection{Von Mises distribution}

The von Mises distribution is a symmetric unimodal distribution, which is given by
$$ \textnormal{VM}(\theta \vert \mu, \kappa) = \{2 \pi I_0(\kappa)\}^{-1} \exp\{\kappa \cos(\theta - \mu)\} , ~~~~~ 0\leq \theta < 2\pi, \kappa \geq 0$$
where $\theta$ represents the data, $\mu$ represents the mean direction, $\kappa$ is the concentration parameter, and $I_0(\cdot)$ is the modified Bessel function of order 0 \citep{abramowitz1972handbook}. A higher $\kappa$ represents less variation, and thus more concentrated data. Let $\boldsymbol\theta=(\theta_1, \dots, \theta_n$) be a sample of angular measurements $\theta_i$ of size $n$.

\begin{figure}
\centering
\includegraphics[width=0.7\linewidth, clip, trim=4cm 3cm 3cm 2cm]{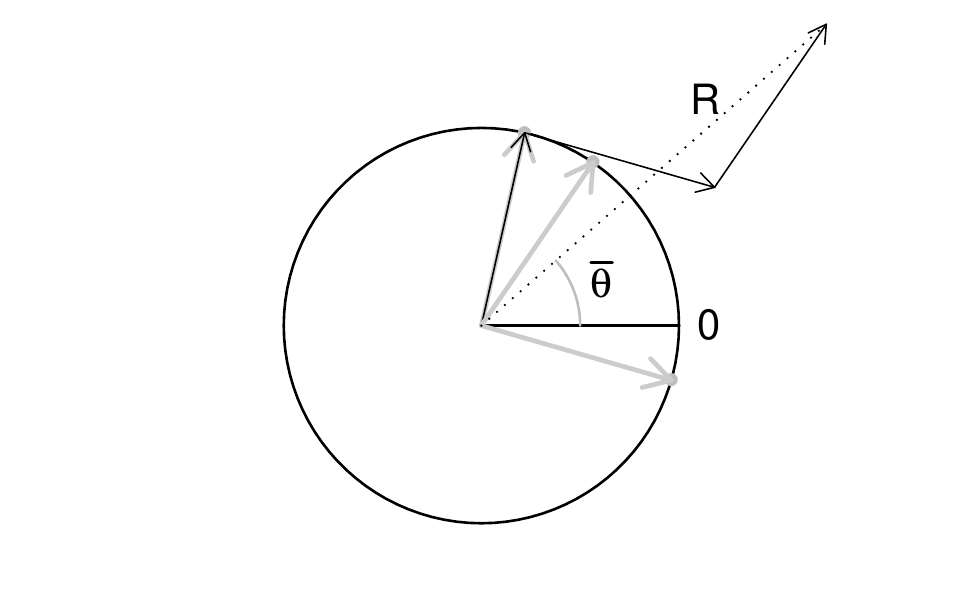}
\caption{Illustration of the mean direction and resultant length of  $\boldsymbol\theta=\{56^\circ, 77^\circ, 344^\circ\}$. The summation of the vectors results in a vector of length $R$ in direction $\bar{\theta}$.}
\label{exampleRMu}
\end{figure}

Each angle in the dataset may be viewed as a vector of length 1 in direction $\theta_i$. As illustrated in Figure \ref{exampleRMu}, the summation of these vectors results in a vector in direction $\bar{\theta}$ of length $R$. $\bar{\theta}$ is an unbiased estimator of $\mu$, while $R$ is called the resultant length and may be obtained from
$$ R = \sqrt{\left(\sum_{i=1}^{n} \cos \theta_i \right)^2 + \left(\sum_{i=1}^{n} \sin \theta_i \right)^2},$$
which increases with concentration and sample size. In the von Mises model, $R$ is a sufficient statistic for $\kappa$. The mean resultant length can be computed as $\bar{R} = R/n$, which is a metric of concentration independent of the sample size.

\subsection{Prior distribution} \label{prior}

\citet{guttorp1988finding} present a conjugate prior for the von Mises distribution. It is given up to a constant of proportionality by 
$$ p(\mu, \kappa) \propto  I_0 (\kappa) ^{-c} \exp\{R_0 \kappa \cos (\mu - \mu_0)\}  ,$$
which represents $c$ observations with prior mean direction $\mu_0$ and prior resultant length $R_0$. In all methods applied in this paper, this conjugate prior will be used. 

\subsection{Posterior distribution} \label{posterior}
To obtain the posterior distribution, the data and the prior are combined to obtain the posterior mean $\mu_n$\footnote{In  R \citep{team2013r}, calculation of $\mu_n$ is readily available in \texttt{atan2($S_n$, $C_n$)}.} and the posterior resultant length $R_n$ by
$$ C_n = R_0 \cos \mu_0 + \sum_{i=1}^n \cos \theta_i, ~~~~ S_n = R_0 \sin \mu_0 + \sum_{i=1}^n \sin \theta_i,$$
$$  \mu_n = \left\{ 
  \begin{array}{l l}
   \tan^{-1} (S_n/C_n)        & \quad \textnormal{if}~ \text{$C_n > 0, S_n > 0$}\\
   \tan^{-1} (S_n/C_n) + \pi  & \quad \textnormal{if}~ \text{$C_n < 0$}\\ 
   \tan^{-1} (S_n/C_n) + 2\pi & \quad \textnormal{if}~ \text{$C_n > 0, S_n < 0$}
  \end{array} \right.$$
and
$$ R_n = \sqrt{C_n^2 + S_n^2}.$$

Then, the joint posterior distribution is given up to a constant of proportionality, by
$$ f(\mu, \kappa \vert \boldsymbol\theta) \propto \{I_0 (\kappa) \}^{-m} \exp\{R_n \kappa \cos (\mu - \mu_n)\}, $$
where $m = n + c$. This distribution is not of closed form due to the Bessel function.

\subsection{Conditional distributions \label{distpar}}

The MCMC methods presented in Section \ref{methods} are based upon the conditional posterior distributions $f(\mu \vert \kappa, \boldsymbol\theta)$ and $f(\kappa \vert \mu, \boldsymbol\theta)$. The conditional posterior distribution of $\mu$, up to a constant of proportionality, is given by
$$f(\mu \vert \kappa, \boldsymbol\theta) \propto \exp\{R_n \kappa \cos(\mu - \mu_n)\},$$
which is the kernel of a von Mises distribution with mean direction $\mu_n$ and concentration parameter $R_n\kappa$. Several straightforward methods to sample data from the von Mises distribution are available.\citep{best1979efficient, fisher1995statistical}

The conditional distribution of $f(\kappa \vert \mu, \boldsymbol\theta)$, is given by $$ f(\kappa \vert \mu, \boldsymbol\theta) \propto \{ I_0(\kappa) \} ^{-m} \exp\{R_n \kappa \cos(\mu - \mu_n)\}. $$
However, it is not straightforward to sample from this conditional distribution, so that special methods are required. In Section \ref{methods}, three methods that can sample the concentration parameter will be discussed. 

\subsection{Notation for multiple groups \label{multiple}}

Here, basic notation and properties will be defined that will be used to extend the methods discussed in Section \ref{methods} to multiple groups. Denote the groups by $j=1, \dots, J$. Then, for group $j$, the posterior mean is denoted by $\mu_{nj}$ and the posterior resultant length by $R_{nj}$. The sample size of group $j$ is denoted by $n_j$, which will be combined with the prior property $c_j$ to obtain $m_j = n_j + c_j$. Finally, let $$ R_t = \sum_{j=1}^{J} R_{nj} ~~\textnormal{and} ~ m_t = \sum_{j=1}^{J} m_j.$$

Utilising this notation, the posterior for multiple groups with a common $\kappa$ is given by $$f(\boldsymbol{\mu}, \kappa \vert \boldsymbol\theta) \propto \{I_0 (\kappa) \}^{-m_t} \exp \left[ \kappa \sum_{j=1}^{J} R_{nj} \cos (\mu_j - \mu_{nj})\right], $$ where $\boldsymbol{\mu} = (\mu_{1}, \dots, \mu_{J})$ denotes the mean directions of the groups.

\section{MCMC Sampling \label{methods}}

In this section, Bayesian methods will be discussed that are able to sample from the posterior of a von Mises distribution in a between-subjects design with $J \geq 1$ independent groups with common but unknown $\kappa$. Specifically, three novel MCMC methods will be presented: a Gibbs sampler based on work by \citet{damien1999fullbayes}, a Metropolis-Hastings sampler, and a rejection sampler based on work by \citet{forbes2014fast}. Importantly, all three methods employ the conjugate prior as described in Section \ref{prior}. 

\subsection{A Gibbs sampler using latent variables  \label{dw}}

In one of the earliest attempts at sampling the concentration parameter of the von Mises distribution, \citet{damien1999fullbayes} provide a Gibbs sampler that only requires sampling of uniform random variates. It is an application of the procedure of adding latent variables to a posterior distribution in order to be able to apply Gibbs samplers in situations where this may not have been feasible originally \citep{damlen1999auxiliary}.

Although the relative simplicity of the Gibbs sampler usually is appealing, it has been noted that this sampler shows high autocorrelation for more concentrated data, causing slow convergence \citep[p. 990]{nunez2005bayesian}.

\citet{damien1999fullbayes} add latent variables $w, v, x,$ and $u=(u_1, u_2, \dots)$ to the joint posterior density $f(\mu, \kappa \vert \boldsymbol\theta)$, where $u$ is an infinite set of latent variables. It is not necessary to sample an infinite number of values for $u_k$, as computing values for $u_k$ up to some sufficient $k$ provides a good approximation of the correct solution. Let $Z$ be the number of values of $u_k$ that will be sampled, so that the set of sampled values is $u_1, \dots, u_Z$. For each analysis performed with this method, a value for $Z$ must be chosen. This is a disadvantage of this method, because setting $Z$ too high will prove computationally intensive, while setting $Z$ too low produces biased results. 

Another disadvantage is that this method requires setting starting values  not only for $\mu$ and $\kappa$, but also for $w$. However, $w$ does not have an intuitive interpretation, making the choice of a starting value somewhat arbitrary and possibly difficult.

\subsubsection{Sampler for a single group}

The posterior density of a single group, after inclusion of the latent variables, is given up to a constant of proportionality as
\begin{multline*}
 f (\mu, \kappa, w, v, u, x \vert \boldsymbol\theta)  \propto e^{-R_n \kappa} I(v < e^{R_{n} \kappa \{1+\cos(\mu - \mu_n)\}}, x < w^{m-1}) \times \\ 
 \left( e^{-w} \prod_{k=1}^{\infty} I(u_k < e^{-w\lambda_k\kappa^{2k}}) \right),
 \end{multline*}
for which the marginal for $(\mu, \kappa)$ is $f(\mu, \kappa \vert \boldsymbol\theta)$, as required. The Gibbs sampler works by drawing a value from the conditional distributions of $x, v, \mu, u_k, w$ and $\kappa$ in sequential order, each conditional on the current other values. Further details, including the required conditional distributions, are found in \citet{damien1999fullbayes} and will not be given here, as the Gibbs sampler for a single group is a special case of the Gibbs sampler described next with $J=1$. 

\subsubsection{Sampler for multiple groups \label{gibbsmulti}}

This section will describe the adapted procedure to implement the Gibbs sampler for multiple groups, so that it will sample from the posterior density $f(\boldsymbol\mu, \kappa, w, v, u, x \vert \boldsymbol\theta)$. It differs in two ways from the sampler provided in \citet{damien1999fullbayes}: first, means for multiple groups and a common $\kappa$ are now sampled, and second, some steps were combined or simplified to facilitate implementation. Notably, the sampling of a set of values $u_1, \dots, u_Z$ is rewritten to sample another set of values $N_1, \dots, N_Z$. The extended Gibbs sampler consists of the following 8 steps:

\begin{enumerate}

\item Set $\boldsymbol\mu, \kappa,$ and $w$ to their starting values.  

\item  Draw a random variate $\tau$ from $U(0, 1)$. \label{firstgibbsstep}

\item  For each group $j$, draw a value for $\mu_j$ from $U(\mu_{nj} - \cos^{-1}g,~ \mu_{nj} + \cos^{-1}g),$ where $$ g=\max\left[-1, \frac{\ln \tau}{R_t \kappa} + \frac{\sum_{j=1}^{J} R_{nj} \{ 1 + \cos (\mu_j - \mu_{nj} ) \} } {R_t} - 1 \right]. $$

\item Calculate $ M = \tilde{w} + E,$ where $\tilde{w}$ is the current value of $w$ and $E$ is a random variate drawn from an exponential distribution with rate $I_0(\kappa) - 1$.

\item  Draw a new value for $w$ from $e^{-w} I(\tilde{w}r^{1/(m-1)} < w < M)$, where $r$ is a uniform random variate from $U(0,1)$. 

\item Compute $ N_k = \kappa (1 + F_k)^{1/(2k)},$ where $F_k$ is an exponential r.v. with rate $\tilde{w}(k!)^{-2} (0.5\kappa )^{2k},$ and $k = 1, \dots, Z$. Set $N = \min N_k$. For advice on setting $Z$, see Section \ref{k}. \label{alg:Nkselection} 

\item  Draw a value for $\kappa$ from $e^{-R_n\kappa} I( \max\{0, v_n\} < \kappa < N),$ where $$ v_n = \frac{\ln \tau}{\sum_{j=1}^{J} R_{nj} \{1+\cos(\mu_j - \mu_{nj}) \} } + \kappa.$$ \label{lastgibbsstep}

\item Repeat steps \ref{firstgibbsstep} - \ref{lastgibbsstep} until a sufficient number of samples have been obtained.

\end{enumerate}

\subsubsection{Choosing $Z$ \label{k}}

In step \ref{alg:Nkselection} of the procedure given above to draw from the conditional density of $\kappa$, a number of samples $N_k$ are generated of which the smallest is retained. However, the number of $N_k$  that should be sampled (here denoted by $Z$) was not discussed in \citet{damien1999fullbayes}. A small simulation study was performed to be able to give guidelines for setting $Z$ when applying this algorithm. 

For each combination of sample sizes $\{5, 30, 100\}$ and concentrations $\{0.1, 1, 4, 8, 16, 32\}$, 100 datasets with $J=1$ and $J=3$ were generated. The Gibbs sampler was then run for 10000 iterations with no lag and a burn-in of 500 on each dataset, with $Z$ set to 40. In each iteration, the index number $k$ of the selected (smallest) value for $N_k$ was saved. This resulted in 100 chains (one for each dataset) of chosen index numbers $k$ of 10000 iterations. Then, the overall maximum value of these chains was taken. If in all these iterations the chosen value never exceeds some number, setting $Z$ to that number or slightly above it will ensure that $Z$ is not too low to produce bias while still retaining some computationally efficiency. 

From the results, given in Table \ref{tab:sufficientk}, it is apparent that a value for $Z$ of about $20$ should be sufficient for the current study. The required $Z$ decreases with higher sample sizes and less concentrated data. It is recommended to investigate sensitivity on $Z$ before application.

\begin{table}[h]
\centering
\caption{Maximum $k$ that was picked out as the smallest value after 10000 iterations of the Gibbs sampler applied to 100 datasets for different sample sizes ($n$), concentration ($\kappa$) and number of groups $(J)$.}
\label{tab:sufficientk}
\begin{tabular}{c@{\hskip 1.0cm}cccc@{\hskip 0.7cm}ccc}
  \toprule 
  & & $J=1$ & & & & $J=3$ & \\ 
   \cmidrule{2-4} \cmidrule{6-8} 
  $\kappa$ & $n=10$ & $n=30$ & $n=100$ & & $n=10$ & $n=30$ & $n=100$ \\ 
  \hline 
 0.1 & 7 & 4 & 3 &  & 5 & 4 & 3 \\ 
  1 & 13 & 7 & 7 &  & 6 & 5 & 6 \\ 
  4 & 17 & 11 & 8 &  & 10 & 9 & 7 \\ 
  8 & 17 & 13 & 9 &  & 10 & 9 & 7 \\ 
  16 & 19 & 14 & 9 &  & 13 & 9 & 9 \\ 
  32 & 19 & 12 & 9 &  & 12 & 9 & 8 \\ 
   \bottomrule 
\end{tabular}
\end{table}

\subsection{A Metropolis-Hastings sampler \label{vmmh}}

Another approach is to employ the Metropolis-Hastings (MH) method \citep{metropolis1953equation, hastings1970monte} to sample from the posterior of a von Mises distribution. MH algorithms are often slower and encounter more autocorrelation and convergence problems than Gibbs samplers, as Gibbs sampling can be seen as a special case of the MH algorithm. However, considering the complicated nature of adding latent variables in the Gibbs sampler described above, an MH method may be advantageous. Another advantage is that the algorithm is reasonably straightforward. On the other hand, this method depends on a proper choice for the proposal density, which may limit its use.

\subsubsection{Sampler for a single group}

To apply the sampler for a single group, samples are needed for a single $\mu$ and $\kappa$. The conditional distribution of the mean direction $\mu$ is known and is easy to sample from using a Gibbs step. The conditional distribution of $\kappa$ is known but difficult to sample from, which will be solved by applying an MH step. 

For the MH step, two main ingredients are required: the posterior from which samples are required, and a proposal density from which it is straightforward to sample. The conditional posterior $f(\kappa \vert \mu, \boldsymbol\theta)$ is given in Section \ref{distpar}. As a non-negative proposal density, the $\chi^2$-distribution will be used. More complex and flexible proposal densities, such as the Gamma distribution, may provide benefits, but for the sake of simplicity only the $\chi^2$-distribution will be considered here. The full algorithm will not be presented here as it is a special case of the sampler described next with $J=1$. 

\subsubsection{Sampler for multiple groups \label{MHmulti}}

The MH sampler for multiple groups may employ the posterior $f(\boldsymbol\mu, \kappa \vert \boldsymbol\theta)$ as given in Section \ref{multiple}. However, in order to prevent underflow issues, the natural logarithm of the posterior is used, which is 
$$ \ln f(\boldsymbol{\mu}, \kappa \vert \boldsymbol\theta) \propto - m_t \ln \left[  I_0(\kappa)\right] +  \kappa \sum_{i=1}^{n} R_{nj} \cos(\mu_{j} - \mu_{nj}). $$ 
Let $\kappa_{cur}$ be the current value of $\kappa$, and $\chi^2(x \vert h)$ be the chi-square distribution with $h$ degrees of freedom. Then, the MH method is given by the following \ref{finalmhstep} steps:

\begin{enumerate}
\item Set $\kappa_{cur}$ to its starting value.
\item For each group $j$, draw a value $\mu_{j}$ from $\textnormal{VM}(\mu_{j} \vert \mu_{nj}, R_n\kappa_{cur})$. \label{beginmhloop}
\item Draw a candidate $\kappa_{can}$ from $\chi^2(\kappa_{can} \vert \kappa_{cur})$. 
\item Calculate the MH ratio as 
\begin{align*}
a = ~& \ln f(\kappa_{can} \vert \boldsymbol{\mu}, \boldsymbol\theta) + \ln  \chi^2(\kappa_{cur} \vert \kappa_{can}) \\
   - & \ln f(\kappa_{cur} \vert \boldsymbol{\mu}, \boldsymbol\theta) - \ln \chi^2(\kappa_{can} \vert \kappa_{cur}),
\end{align*}
where $\boldsymbol\mu = \lbrace \mu_1, \dots, \mu_J \rbrace ,$ a row vector of current values of $\mu$ for each group.
\item Draw a value $u$ from $U(0,1)$. 
\item If $a > \ln u $, set $\kappa_{cur} = \kappa_{can}$. Elsewise, remain at $\kappa_{cur}$. \label{endmhloop}
\item Repeat step \ref{beginmhloop} - \ref{endmhloop} until a sufficient number of samples have been obtained. \label{finalmhstep}
\end{enumerate}

\subsection{A rejection sampler \label{fm}}

In a recent paper, \citet{forbes2014fast} present a promising new algorithm to sample from the conditional posterior $f(\kappa \vert \mu, \boldsymbol\theta)$. The approach is largely focused on computational speed, and was motivated by the fact that plugging a Bessel function approximation into the von Mises posterior leads to a Gamma distribution.

The approach performs rejection sampling for $\kappa$ on the basis of two parameters, $\{\eta, \beta_0\}$. For a single group of data with no prior, as described in \citet{forbes2014fast}, the algorithm sets $\eta = n$ and $\beta_0 = - n^{-1} \sum_{i=1}^{n} \cos(\theta_i - \mu)$. These are then used to compute the approximately optimal parameters for a Gamma proposal, such that the probability of rejection is minimized. In the rejection step, a candidate for $\kappa$ is then repeatedly drawn from this Gamma proposal density until it is accepted. 

Samples of $\mu$ are drawn outside of the algorithm, which may be done easily as in step \ref{beginmhloop} of the MH procedure in Section \ref{MHmulti}. As with the MH method, only $\kappa$ requires a starting value. 

\subsubsection{Sampler for a single group}

\citet{forbes2014fast} describe the rejection sampler for a single group of data using a constant prior. The conjugate prior that is preferred here can be added as described below. 

Using the sample mean direction $\bar{\theta}$, it can be shown that
$$ \beta_0 = - n^{-1} \sum_{i=1}^{n} \cos(\theta_i - \mu) = - \frac{R\cos (\mu - \bar\theta) }{n}.$$
This relation to the resultant length means that $\mu_n, R_n,$ and $m$ from the desired posterior can be plugged into the formula for $\beta_0,$ to obtain
$$\beta_n = - \frac{R_{n} \cos (\mu - \mu_{n})}{m}.$$
Then, the rejection algorithm can be applied exactly as given in  \citet{forbes2014fast}, using $\beta_n$ instead of $\beta_0$ and $\eta = m$.

\subsubsection{Sampler for multiple groups}

As the sampling of mean directions occurs outside of the main algorithm, it is straightforward to sample separate mean directions for each group. However, the common $\kappa$ depends on the sampled means through $\beta_n$. After computation of $\beta_n$, the rejection algorithm no longer uses the data $\boldsymbol\theta$ or the current value of $\mu$. The sampler will thus be extended to multiple groups by once again rewriting $\beta_n$. 

Using $R_{nj}$ and $\mu_{nj}$, and $m_t$ as before, let
$$\beta_t = -  \frac{\sum_{j=1}^{J} R_{nj} \cos (\mu - \mu_{nj})}{m_t}.$$
Then, the rejection algorithm can be applied using $\beta_t$ instead of $\beta_0$ and $\eta = m_t$.

\section{Simulation study \label{simstud}}

In the previous section, three distinct methods to sample from the posterior of the von Mises distribution with multiple groups were shown. In this section, these three methods will be evaluated on their performance and efficiency. 

\subsection{Methods}

\begin{figure}[bt]
\centering
\includegraphics[width=\textwidth]{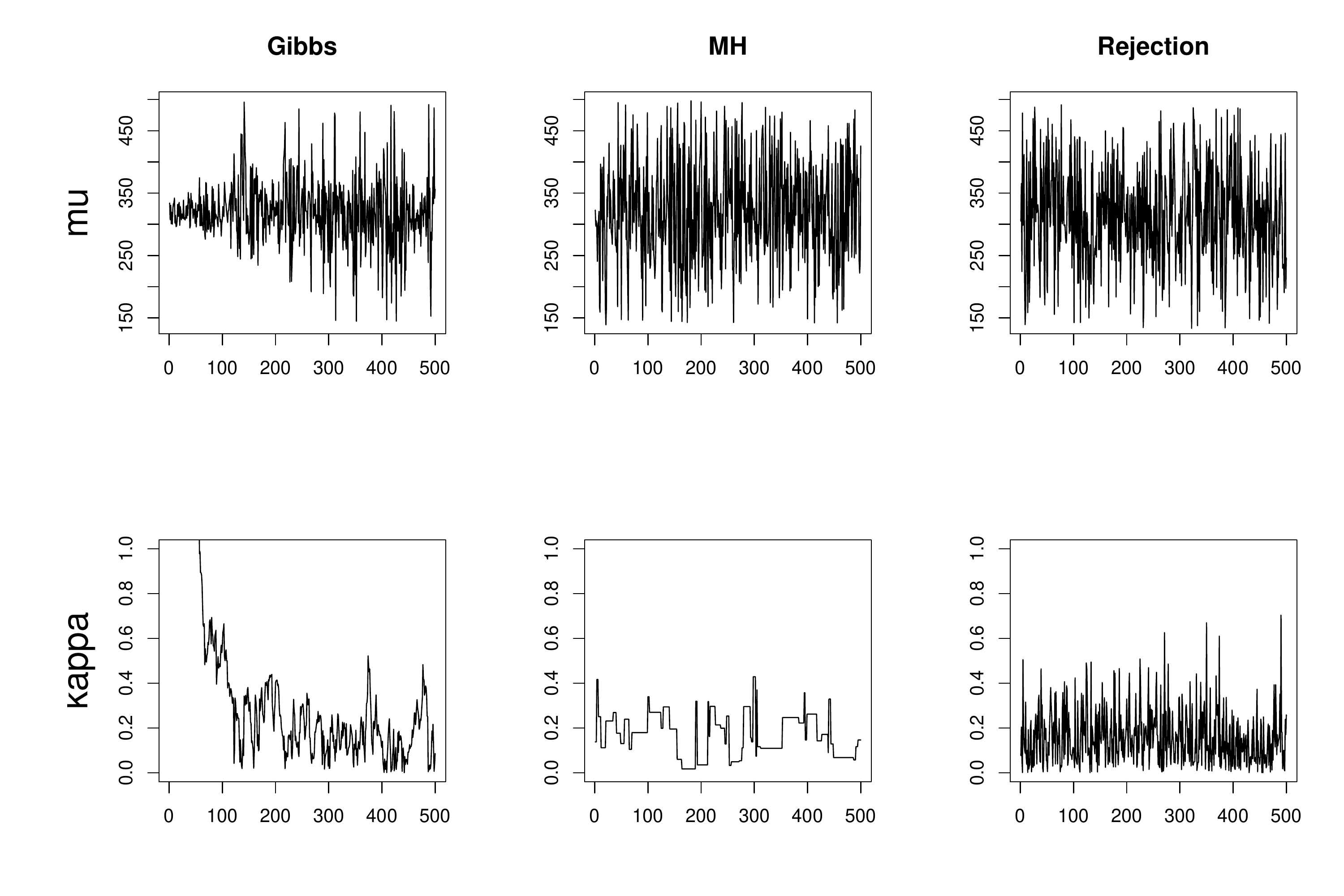}
\caption{Example chains of $\mu$ (in degrees) and $\kappa$ drawn in the first 500 iterations of each of the three methods, with no burn-in and without thinning the chain, where $J = 3$, true $\kappa = 0.1,$ and $n_j = 30$.}
\label{example}
\end{figure}

All three methods were implemented in C++ within R \citep{team2013r} via Rcpp \citep{rcpp}. To illustrate the differences between the three methods, Figure \ref{example} shows example chains of the first 500 iterations for each of the three methods. It can be seen that the Gibbs sampler has large autocorrelation and slow convergence, that the MH algorithm can have low acceptance probability but converges fast, and that the rejection algorithm converges fast and mixes well. 

The three sampling methods were applied to various scenarios, which differed in the following three properties. First, the samplers analyzed both a single group of data ($J=1$) and three groups of data ($J=3$). Second, sample sizes of 10, 30 and 100 were used. For $J=3$, this sample size denotes the sample size per group ($n_j$), making the total sample size $3n_j$. Third, values for the concentration parameter $\kappa$ were 0.1, 4 and 32. Because the multiple groups are assumed have equal $\kappa$, all three groups of data were sampled given the same true $\kappa$. These manipulations resulted in a 3x2x3x3 simulation study design, for a total of 54 cells. For $J=1$, the true mean was set at $20^\circ$, while true means for $J=3$ were set at $20^\circ, 40^\circ,$ and $60^\circ$.

For each cell, 2000 datasets were generated, each of which was analyzed with each sampler. Burn-in and lag (that is, how much the chain will be thinned) were set to appropriate values (see Section \ref{convergence}), after which the first 10000 retained iterations of both $\boldsymbol\mu$ and $\kappa$ were saved. Although all three methods allow inclusion of prior information through the conjugate prior, a non-informative prior was used throughout the simulation study by setting $\mu_0=0, R_0=0,$ and $c=0$. Each method requires a starting value for $\kappa$, which was set to 2 in all cases. The Gibbs sampler required additional starting values for $\mu$ and $w$, which were set at 0 and 4 respectively, regardless of sample size or $\kappa$. Additionally, for the Gibbs sampler an appropriate $Z$ must be chosen (see Section \ref{k}), which was set to $25$ throughout this study. 

\subsection{Convergence} \label{convergence}

As convergence is achieved at a different number of iterations for each of the methods, several runs of each were assessed for each cell in order to assess convergence and required burn-in and lag, which were then set correctly for the simulation study. Burn-in was set, in all cases, to 500 times the chosen value for the lag. 

The Gibbs sampler performed adequately for small samples with large dispersion. For example, a single group of 10 datapoints with true $\kappa = 0.1$ produced a reasonable sample from the posterior using a lag of 2, which means saving every other iteration. With larger sample sizes and more concentrated data, the autocorrelation increases quickly, requiring a lag of 250 for $\kappa = 4$ and $n_j = 100$ with $J=3$. For values of $\kappa$ above about 7, application of the Gibbs sampler becomes unfeasible, so results for the Gibbs sampling method with true $\kappa=32$ are not reported. 

The MH algorithm fared much better, converging quickly in all tested situations. However, applying MH methods requires reasonable acceptance rates, which can be computed by $Q_{acc}/Q$, where $Q_{acc}$ is the number of accepted iterations, and $Q$ is the total number of iterations.  \citet{johnson1999ordinal} suggest an acceptance rate of about 50\% to be ideal. A low acceptance rate may suggest a badly fitting proposal density, while a high acceptance rate (i.e. close to 1) may suggest that the algorithm has yet to converge properly. As convergence was assessed seperately and achieved quite quickly, only too low acceptance rates were of concern here. Acceptance rates increased for smaller sample sizes and more concentrated data. For example, for a single group of data with $n=100$ and $ \kappa=0.1$ the average acceptance rate was as low as 0.1, while for $n=10, \kappa=32$ the acceptance rate was as high as .81. 

The rejection algorithm converged almost immediately and showed barely any autocorrelation. An acceptance rate can be computed by $Q/Q_{can}$, where $Q$ is the total number of accepted candidates, which was chosen beforehand as the desired number of iterations, and $Q_{can}$ is the total number of candidates, including those that were accepted. The algorithm rejected no more than 15\% of the candidates in any case.

\begin{table}
\vspace{-0.76cm}
\begin{minipage}{1.06\textwidth}
\begin{center}
\caption{\label{J1}Average of posterior properties over 2000 replications for $J=1$, with true $\mu = 20^\circ$, for different sample sizes $(n)$ and concentration $\kappa$.}
{\begin{tabular}{ccccccccc}
  \toprule 
 &&& \multicolumn{2}{c}{Posterior $\mu$} & \multicolumn{2}{c}{Posterior $\kappa$\footnote{Posterior $\kappa$ mode denotes the mode as described in section \ref{hdimode}. Coverage denotes the proportion of replications in which the true $\kappa$ fell within the 95 \% HDI.}}&  \\
   n & $\kappa$ & Method & Mean & Coverage & Mode & Coverage
                                            & Acc.\footnote{Acceptance ratio. For Gibbs sampling, this is always 1. For Metropolis-Hastings, $Q_{acc}/Q$ is given. For the rejection method, this is $Q/Q_{can}$.} & MCT\footnote{Mean Computation Time of one replication in seconds.} \\
 \midrule 
10 & 0.1 & Gibbs & 15.52 & 0.74 & 0.34 & 0.98 & 1 & 0.74 \\ 
   &  & MH & 15.72 & 0.76 & 0.34 & 0.96 & 0.24 & 0.02 \\ 
   \vspace{0.2cm} &  & Rejection & 15.70 & 0.75 & 0.36 & 0.97 & 0.91 & 0.02 \\ 
   & 4 & Gibbs & 20.26 & 0.97 & 4.73 & 0.93 & 1 & 9.60 \\ 
   &  & MH & 20.26 & 0.96 & 5.03 & 0.94 & 0.53 & 0.02 \\ 
   \vspace{0.2cm} &  & Rejection & 20.26 & 0.96 & 4.90 & 0.96 & 1 & 0.02 \\ 
   & 32 & Gibbs & --- & --- & --- & --- & --- & --- \\ 
   &  & MH & 19.91 & 0.92 & 40.94 & 0.95 & 0.81 & 0.02 \\ 
   &  & Rejection & 19.91 & 0.92 & 41.42 & 0.95 & 1 & 0.02 \\ 
   \cmidrule{1-9}30 & 0.1 & Gibbs & 22.35 & 0.81 & 0.19 & 0.98 & 1 & 1.08 \\ 
   &  & MH & 22.03 & 0.81 & 0.18 & 0.96 & 0.16 & 0.02 \\ 
   \vspace{0.2cm} &  & Rejection & 22.15 & 0.80 & 0.20 & 0.97 & 0.89 & 0.02 \\ 
   & 4 & Gibbs & 19.99 & 0.97 & 4.20 & 0.94 & 1 & 11.50 \\ 
   &  & MH & 19.99 & 0.97 & 4.27 & 0.96 & 0.36 & 0.02 \\ 
   \vspace{0.2cm} &  & Rejection & 19.99 & 0.97 & 4.18 & 0.97 & 1 & 0.02 \\ 
   & 32 & Gibbs & --- & --- & --- & --- & --- & --- \\ 
   &  & MH & 19.98 & 0.94 & 34.43 & 0.95 & 0.70 & 0.02 \\ 
   &  & Rejection & 19.98 & 0.94 & 34.40 & 0.95 & 1 & 0.02 \\ 
   \cmidrule{1-9}100 & 0.1 & Gibbs & 20.82 & 0.86 & 0.11 & 0.99 & 1 & 3.58 \\ 
   &  & MH & 20.97 & 0.87 & 0.11 & 0.98 & 0.10 & 0.02 \\ 
   \vspace{0.2cm} &  & Rejection & 20.91 & 0.86 & 0.12 & 0.98 & 0.86 & 0.02 \\ 
   & 4 & Gibbs & 19.99 & 0.96 & 4.06 & 0.94 & 1 & 57.59 \\ 
   &  & MH & 19.99 & 0.95 & 4.08 & 0.94 & 0.21 & 0.02 \\ 
   \vspace{0.2cm} &  & Rejection & 19.99 & 0.96 & 4.02 & 0.96 & 1 & 0.02 \\ 
   & 32 & Gibbs & --- & --- & --- & --- & --- & --- \\ 
   &  & MH & 20.03 & 0.95 & 32.72 & 0.96 & 0.53 & 0.02 \\ 
   &  & Rejection & 20.03 & 0.95 & 32.72 & 0.96 & 1 & 0.02 \\ 
   \bottomrule 
\end{tabular}
}
\end{center}

\vspace{-0.76cm}

\end{minipage}
\end{table}

\subsection{Mode estimation for $\kappa$ \label{hdimode}}

Estimating $\kappa$ as the mean or the median of the posterior sample may lead to biased results, as $\kappa$ is non-negative and has a right-skewed distribution. For skewed distributions, the mode usually provides the least biased estimate. An estimate of the mode can be obtained by using the Highest Density Interval (HDI), which is the shortest interval containing a certain percentage of the data \citep{venter1967estimation}. Here, the mode was estimated to be the midpoint of the 10\% HDI. 

\subsection{Results}

In Tables \ref{J1} and \ref{J3}, results are displayed for a single group and three groups, respectively. As mentioned before, applying the Gibbs sampler to a situation with $\kappa=32$ is unfeasible, and therefore these rows are left empty. 

The column below posterior $\mu$ mean gives the average of the posterior mean direction of either $\mu$ or $\{\mu_1, \mu_2, \mu_3 \}$ of all replications. The coverage of the mean denotes the proportion of replications where 95\% Central Credible Interval (CCI) contained the true $\mu$. For $J=3$, this coverage was averaged over the three means. The desired value of the coverage is .95. For the posterior $\kappa$, the estimated mode for each replication was saved, as well as the 95\% HDI. The average of the mode over replications is provided in the column posterior $\kappa$ mode, followed by the posterior $\kappa$ coverage, which denotes the proportion of replications for which the true value fell within the 95\% HDI. The last two columns provide the acceptance rate and the mean computation time (MCT) per replication in seconds. 

The size of the bias tends to depend on the true value. In order to investigate by what factor estimates are off, a relative bias can be calculated as $Bias/True~value$. The relative bias may help facilitate interpretation of relative severity of bias for $\kappa$, in order to allow for more accurate comparisons between cells. 

\subsubsection{A single group}
\paragraph{Posterior $\mu$}

All three methods provided similar results for the posterior mean, which was generally close to the true mean. Estimates were closer to the true value for increasing $n$ and increasing $\kappa$. The worst case was found for $\kappa=0.1, n=10$, where the difference between the true $\mu$ ($20^\circ$) and average posterior $\mu$ was about $4.3^\circ$. However, this difference can almost surely be attributed to sampling error of datasets instead of an issue with the MCMC methods. When $\kappa=0.1$, the distribution of the sample mean direction $\bar{\theta}$ is close to the circular uniform distribution, so that the average over the sample mean directions, even over 2000 datasets, shows some random variation. This is supported by the fact that the MCMC methods all show the same difference from the true value. In general, there seems to be no systematic bias in the estimation of the mean direction. 

Coverage was generally adequate as well. Undercoverage was observed for $\kappa = 0.1$ with all sample sizes, although the coverage improved with increasing sample size. Coverage for $\mu$ relies on correct procedures of sampling both $\mu$ and $\kappa$. For example, an upwards bias in $\kappa$ results in a lower coverage for $\mu$. Sampling methods for $\mu$ are straightforward, and thus deviations of the coverage from .95 are likely due to a deficient mechanism to sample $\kappa$, as the current value of $\kappa$ is used in the distribution of $\mu$.

\paragraph{Posterior $\kappa$}

The mode of $\kappa$ shows a systematic upward bias for all cells and all methods. The relative bias is worse for smaller $\kappa$ and thus more dispersed data. The bias also decreases with increasing $n$ and nearly disappears for $n = 100$. This bias coincides with a well-known bias in maximum likelihood estimation of $\kappa$ \citep[p. 87]{mardia1999directional}. For suggestions on corrections to obtain unbiased estimates, see \citet{best1981bias}.

Regardless of the observed bias, coverage for $\kappa$ was generally acceptable, fluctuating around .95 for all methods. In conditions with low concentration, a tendency towards overcoverage (coverage above .95) can be seen.

\paragraph{Mean Computation Time}

The final column denotes the computational time for the algorithms as implemented in C++ via Rcpp, which was averaged over all replications. The Gibbs sampling method performed worst by far. Its computational time increased with higher sample sizes and higher concentration. The longest reported time was 57.59 seconds per replication. Both the MH and rejection algorithm were very fast. Their computational time was fairly independent of both sample size and $\kappa$, so that it always took about 0.02 seconds for a replication.

\subsubsection{Multiple groups}

\begin{table}
\centering
\begin{minipage}{1.06\textwidth}
\caption{Average of posterior properties over 2000 replications for $J=3$, with true means $\mu_1 = 20^\circ$, $\mu_2 = 40^\circ$, $\mu_3 = 60^\circ$, for different samples sizes per group $(n_j)$ and concentration $(\kappa)$.}
\label{J3}

{\footnotesize
\begin{tabular}{ccccccccccc}
  \toprule 
 &&& \multicolumn{4}{c}{Posterior $\mu$} & \multicolumn{2}{c}{Posterior $\kappa$\footnote{Posterior $\kappa$ mode denotes the mode as described in section \ref{hdimode}. Coverage denotes the proportion of replications in which the true $\kappa$ fell within the 95 \% HDI.}
 }&  \\
   $n_j$ & $\kappa$ & Method & $\mu_1$ & $\mu_2$ & $\mu_3$ & Coverage & Mode & Coverage  & Acc.\footnote{Acceptance ratio. For Gibbs sampling, this is always 1. For Metropolis-Hastings, $Q_{acc}/Q$ is given. For the rejection method, this is $Q/Q_{can}$.} & MCT\footnote{Mean Computation Time of one replication in seconds.}  \\
 \midrule
 10 & 0.1 & Gibbs & 21.55 & 42.64 & 64.59 & 0.95 & 0.07 & 1 & 1 & 3.32 \\ 
   &  & MH & 21.99 & 43.10 & 65.30 & 0.92 & 0.23 & 0.97 & 0.17 & 0.05 \\ 
   \vspace{0.2cm} &  & Rejection & 22.17 & 43.11 & 65.07 & 0.91 & 0.26 & 0.98 & 0.91 & 0.05 \\ 
   & 4 & Gibbs & 20.01 & 40.11 & 60.02 & 0.97 & 4.06 & 0.94 & 1 & 7.47 \\ 
   &  & MH & 19.99 & 40.14 & 60.01 & 0.96 & 4.35 & 0.95 & 0.36 & 0.05 \\ 
   \vspace{0.2cm} &  & Rejection & 19.99 & 40.13 & 60.02 & 0.96 & 4.26 & 0.96 & 1 & 0.05 \\ 
   & 32 & Gibbs & --- & --- & --- & --- & --- & --- & --- & --- \\ 
   &  & MH & 19.98 & 39.93 & 60.10 & 0.94 & 34.51 & 0.95 & 0.70 & 0.05 \\ 
   &  & Rejection & 19.98 & 39.93 & 60.10 & 0.94 & 34.55 & 0.96 & 1 & 0.05 \\ 
   \cmidrule{1-11} 
 30 & 0.1 & Gibbs & 15.16 & 39.28 & 51.20 & 0.97 & 0.04 & 1 & 1 & 1.30 \\ 
   &  & MH & 15.42 & 40.69 & 50.34 & 0.94 & 0.14 & 0.98 & 0.11 & 0.05 \\ 
   \vspace{0.2cm} &  & Rejection & 15.68 & 40.68 & 50.36 & 0.94 & 0.15 & 0.98 & 0.88 & 0.05 \\ 
   & 4 & Gibbs & 19.97 & 40.09 & 59.86 & 0.96 & 3.99 & 0.93 & 1 & 8.66 \\ 
   &  & MH & 19.95 & 40.09 & 59.83 & 0.96 & 4.07 & 0.95 & 0.22 & 0.05 \\ 
   \vspace{0.2cm} &  & Rejection & 19.96 & 40.10 & 59.83 & 0.96 & 4.01 & 0.96 & 1 & 0.05 \\ 
   & 32 & Gibbs & --- & --- & --- & --- & --- & --- & --- & --- \\ 
   &  & MH & 19.94 & 40.03 & 60.02 & 0.94 & 32.84 & 0.95 & 0.55 & 0.05 \\ 
   &  & Rejection & 19.94 & 40.03 & 60.02 & 0.95 & 32.79 & 0.95 & 1 & 0.05 \\ 
   \cmidrule{1-11} 
 100 & 0.1 & Gibbs & 18.00 & 39.06 & 58.53 & 0.97 & 0.04 & 1 & 1 & 4.77 \\ 
   &  & MH & 18.08 & 39.41 & 58.42 & 0.95 & 0.10 & 0.97 & 0.07 & 0.05 \\ 
   \vspace{0.2cm} &  & Rejection & 18.30 & 39.49 & 58.57 & 0.95 & 0.10 & 0.98 & 0.86 & 0.05 \\ 
   & 4 & Gibbs & 19.98 & 39.98 & 60.10 & 0.95 & 4.01 & 0.93 & 1 & 31.21 \\ 
   &  & MH & 19.96 & 39.98 & 60.10 & 0.95 & 4.04 & 0.94 & 0.13 & 0.05 \\ 
   \vspace{0.2cm} &  & Rejection & 19.96 & 39.98 & 60.10 & 0.95 & 3.98 & 0.96 & 1 & 0.05 \\ 
   & 32 & Gibbs & --- & --- & --- & --- & --- & --- & --- & --- \\ 
   &  & MH & 19.98 & 40 & 60.03 & 0.95 & 32.24 & 0.95 & 0.36 & 0.05 \\ 
   &  & Rejection & 19.98 & 40 & 60.03 & 0.95 & 32.21 & 0.95 & 1 & 0.05 \\ 
   \bottomrule 
\end{tabular}
}

\vspace{-0.3cm}
\end{minipage}
\end{table}

In Table \ref{J3}, results for analysis of multiple groups of data are shown. For the posterior group mean directions the observed pattern was similar to the single group case. The Gibbs sampler showed slight overcoverage for $\mu$ when $\kappa=0.1$, while the MH and rejection method show slight undercoverage with $n_j=10, \kappa = 0.1$. In all other cells, coverage of $\mu$ was adequate.

\begin{figure}[bt]
\includegraphics[width=\textwidth]{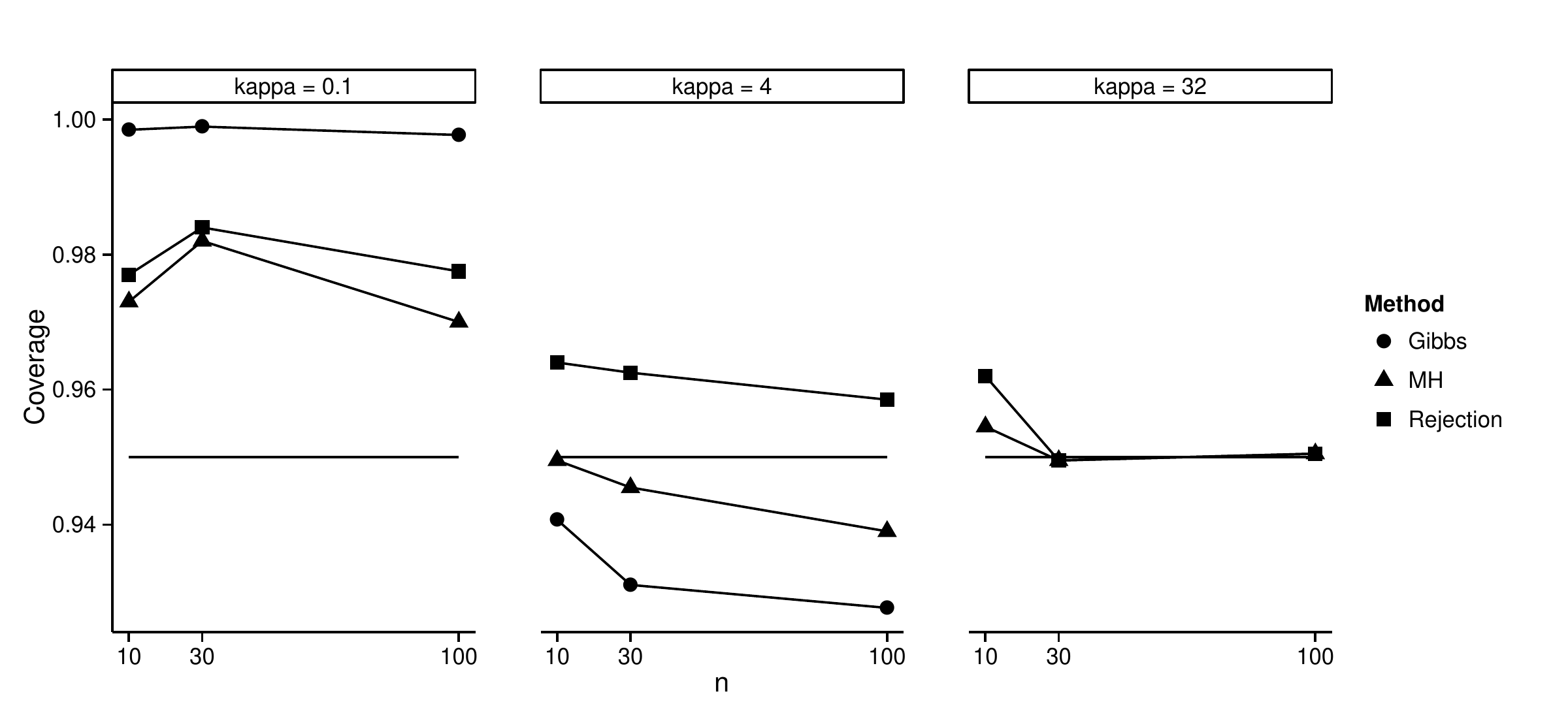}
\caption{Coverages of $\kappa$ for different sample sizes $(n)$ and concentration $(\kappa)$, all with $J=3$. The solid straight line indicates the target of the coverages, .95.}
\label{coverage}
\end{figure}

As with $J=1$, a systematic upward bias was observed in $\kappa$ for the MH and rejection sampler. The bias was generally smaller for $J=3$ compared to $J=1$, because the total amount of observations is three times as large. The strongest bias was observed for $n_j=10$. For $n_j=30$ and $n_j=100$ much less bias was observed. The Gibbs sampler, however, shows a downward bias in these cases. Figure \ref{coverage} shows the coverages of $\kappa$ per method for different sample sizes and concentration, with $J=3$. Coverages for $\kappa$ were mostly adequate, although the Gibbs sampler performed badly, with severe overcoverage with $\kappa=.1$.  The MH and rejection sampler performed well, although the rejection method seemed slightly more prone to overcoverage. These coverages in the range .95-1 indicate that the HDI would be chosen too wide so that the true value falls within the HDI more often than expected. Finally, computational time increased slightly with three groups for MH and rejection.

\section{Discussion \label{discussion}}

This paper presented three different MCMC approaches for Bayesian estimation of the mean directions $\mu_j$ of multiple groups of circular data with common but unknown concentration $\kappa$. These approaches were based on existing knowledge on Bayesian analysis of circular data that could be used for analysis of a single group of circular data. Additionally, a systematic investigation of the performance of the three approaches was performed. 

Comparing the methods, clear differences became apparent. The Gibbs sampler encountered many problems, among which were undesirable coverages, sizable computational time, and complexity in application. The MH method performed adequately, but it does not show desirable acceptance rates for large datasets with small concentration when using the current $\chi^2$ proposal density. The rejection algorithm by \citet{forbes2014fast} was found to be the most promising of the MCMC-methods available in the literature at present, due to fast computational speed, fast convergence and adequate coverage.

The model developed here is still limited in terms of scope; it provides a basic between-subjects design for multiple groups of circular data, but extensions of this model such as a between-within-design or the inclusion of covariates have yet to be developed. Although in the present study the rejection algorithm was the most advantageous, a general MH algorithm may prove more flexible for such extended models due to its more direct approach. It is expected that extending the MCMC methods provided here to more complex models will exacerbate any issues regarding acceptance rates in different ways, so it remains to be seen which method will perform best after such an extension. 

This study is also limited to the assumption that the data follows the von Mises distribution. Much of the literature on circular statistics assumes that circular data encountered in practice will follow this distribution, but this is not always the case. Groundwork for a general method for any kind of circular distribution was provided by \citet{Bhattacharya2009} and employs importance sampling. Because importance sampling relies on defining an additional density to approximate normalizing constants, simpler methods such as the ones presented here are preferred where possible. 

Finally, the question remains whether the assumption of a common $\kappa$ across groups is a reasonable assumption. Circular data to be analyzed must be tested on this assumption. If it does not hold, methods presented in this paper may simply be applied to each group separately, each with $J=1$. 

In sum, the intrinsic approach offers a promising and flexible approach to Bayesian analysis of circular data, and its extension to a model with $J$ multiple groups is an important first step towards developing flexible modeling of circular data in between-subjects designs. 

\section{Remarks}
This work was supported by a Vidi grant awarded to I. Klugkist from NWO, the Dutch Organization for Scientific Research (NWO 452-12-010).

R Code for application of the methods in this article, as well as supplemental files, are available online at \newline \url{github.com/keesmulder/BayesianMultigroupCircularData}.



\bibliographystyle{apa} 
\bibliography{CircularData}


\end{document}